\documentclass{article}

\usepackage{amsmath, amsthm, amssymb, geometry, mathtools, epsfig}
\usepackage[colorlinks,citecolor=green,linkcolor=blue]{hyperref}

\def\G{\Gamma}
\def\no{\nonumber}

\def\dis{\displaystyle}
\def\le{\left(}
\def\ri{\right)}

\begin{document}

\begin{titlepage}
\vskip 1cm
\begin{center}
{\Large \bf Analytical solution to DGLAP integro-differential equation \\ 
\vskip 3mm
via complex maps in domains of contour integrals}   \\
\vskip 10mm  
Gustavo \'Alvarez $^{(a)}$ and Igor Kondrashuk $^{(b)}$ 
\vskip 5mm  
{\it  (a) II. Institut f\"ur Theoretische Physik, Universit\"at Hamburg,  \\ Luruper Chaussee 149,  22761 Hamburg, Germany} \\ 
{\it  (b) Grupo de Matem\'atica Aplicada {\rm \&} Grupo de F\'isica de Altas Energ\'ias, \\ 
          Departamento de Ciencias B\'asicas, Universidad del B\'io-B\'io,  Campus Fernando May, \\
          Av. Andres Bello 720, Casilla 447, Chill\'an, Chile}  
\end{center}

\begin{abstract}
A simple model for QCD dynamics in which the DGLAP integro-differential equation may be solved analytically has been considered in our previous papers arXiv:1611.08787 [hep-ph] and arXiv:1906.07924 [hep-ph].
When such a model contains only one term in the splitting function of the dominant parton distribution, then  Bessel function appears to be  
the solution to this simplified DGLAP equation. To our knowledge, this model with only one term in the splitting function   
for the first time has been proposed by Bl\"umlein in hep-ph/9506403.  In arXiv:1906.07924 [hep-ph] we have shown that a dual integro-differential equation obtained from 
the DGLAP equation by a complex map in the plane of the Mellin moment in this model may be considered as the BFKL equation. 
Then, in  arXiv:1906.07924 we have applied a complex diffeomorphism to obtain a standard integral  from  Gradshteyn and Ryzhik tables 
starting from the contour integral for parton distribution functions that is usually taken by calculus of residues.  
This standard integral from these tables appears to be  the Laplace transformation of  Jacobian for this complex diffeomorphism.  
Here we write up all the formulae behind this trick in detail and find out certain important points for further development of this strategy. 
We verify that the inverse Laplace transformation of the Laplace image of the Bessel function may be represented in a form of  Barnes contour integral.   
\vskip 0.5 cm
\noindent Keywords: DGLAP equation, complex maps, Jacobians  
\vskip 0.5 cm
\end{abstract}
\end{titlepage}

\section{Introduction}

It often happens  that a solution to an integro-differential equation is obtained in a form of contour integrals in one or more complex planes. 
Such integrals may be taken via Cauchy integral formula by calculating residues. Usually they are not the classical Barnes integrals which are a convenient 
representation for  generalized hypergeometric functions. The integrands of the Barnes integrals are ratios of certain products of several Euler Gamma functions. 
A transformation of a contour integral representing a solution to an integro-differential equation to this form of the Barnes integrals  would be helpful  because they may be  classified 
in terms of  suitable special functions. To find such a transformation is the aim of this paper. The intermediate step will be a representation of  these integrals as the Laplace transformations of the Jacobians 
of some complex maps in the complex variable plane of the contour integral. We have considered such a possibility in the previous paper \cite{Kondrashuk:2019cwi} in which we transformed the contour integral  
representing the solution to the DGLAP integro-differential equation in a simple model of QCD dynamics from this obtained form of a contour integral in the complex  plane of the Mellin moment 
to the Laplace transformation of the corresponding Jacobian. These Jacobians may appear to be  multivalued functions of new complex variables and integration over cuts may be required. 
We may avoid integration over these cuts if represent the integrals with multivalued Jacobians obtained by the first complex map in the form of the Barnes integrals 
by applying one complex map more.  The holomorphic maps of  variables in the complex domains of the contour integrals which we apply in the present paper are based 
on the standard theory of complex variable which may be found in any textbook on this subject \cite{Churchill}. As to  manipulations with closed contours in the complex plane of the integration variable, 
they were already used  
in quantum field theory when integral transformations are involved in calculations \cite{Kosower:1997vj}. We rectify or curve the integration contours too when it is necessary here.

The contour integral we consider in this  paper is a solution to the DGLAP integro-differential equation in a simple model of QCD dynamics considered in Ref. \cite{Alvarez:2016juq}. 
The DGLAP equation was written in the seventies  for the structure functions of proton. They may be measured experimentally 
in the deep inelastic scattering processes \cite{Bjorken:1968dy}. In Refs. \cite{Gribov:1972ri,Gribov:1972rt,Lipatov:1974qm} Gribov and Lipatov studied these processes in QED and  found that these structure functions 
satisfy certain integro-differential equations. The discovery of QCD  has been marked by the Nobel prize paper \cite{Gross:1974cs} in which the  renormalization group equations for the Mellin moments of the coefficient functions 
of Wilson operator product expansion for the matrix element of two currents of the deep inelastic scattering process  have been obtained.   
Then, in Ref. \cite{Altarelli:1977zs} Altarelli and Parisi wrote these renormalization group equations for the coefficient functions 
of the operator product expansion in an integro-differential form in the space of Bjorken $x$ and interpreted them as integro-differential equations for the parton distribution functions.  
In Ref.  \cite{Dokshitzer:1977sg}  Dokshitzer developed to the QCD case the Gribov and Lipatov approach used in QED and wrote integro-differential equations similar to Altarelli-Parisi equations.
These integro-differential equations became known as Dokshitzer-Gribov-Lipatov-Altarelli-Parisi equation (also known as DGLAP equation).

The splitting functions are the input in the DGLAP equation. They may be found from anomalous dimensions of operators in QCD \cite{Vogt:2004mw,Moch:2004pa} and are some combinations of several terms 
\cite{Vogt:2004mw,Moch:2004pa}. Residue calculus via the Cauchy integral formula for the contour integral in the complex plane of the Mellin moment which represents a solution to the DGLAP 
equation is straightforward but this calculus is not simple in the real world because many infinite sums are involved in the result and these sums should be classified \cite{Ablinger:2010kw,Ablinger:2013hcp}.
However, a simple model with only one term in the splitting function may be considered and the DGLAP equation in this case may be solved in terms of the Bessel function.  
To our knowledge, this simplification to one Bessel function of the solution to the DGLAP equation in the case of only one term in the splitting function  for the first time has been mentioned in 1995 in Ref.\cite{Blumlein:1995eu}.  
We have considered this model in detail in our papers \cite{Alvarez:2016juq} and  \cite{Kondrashuk:2019cwi}.
In this paper we transform this contour integral of the simple model of \cite{Blumlein:1995eu,Alvarez:2016juq,Kondrashuk:2019cwi} via a complex map to the form of the Laplace transformation of the Jacobian
of the corresponding complex diffeomorphism and then to the form of the Barnes integrals via another complex map. As the result the contour integral in the plane of the Mellin moment transforms 
to another contour integral in some complex domain.  The integrand after these consequent complex maps transforms to a ratio of certain Gamma functions 
which is a typical form of the Barnes integrals.

We should defend usefulness of the proposed strategy for the DGLAP community because  there is already a long history 
of many achievements related to this equation.
Indeed, the DGLAP integro-differential equation may be converted to a first-order differential equation by 
taking the Mellin moment of both the sides of the DGLAP equation with respect to Bjorken variable $x$ \cite{Altarelli:1977zs}. 
The resulting differential equation is the renormalization group equation 
for the Mellin moments \cite{Altarelli:1977zs} with respect to the scale of momentum transfer in the process of deep inelastic scattering. These differential equations may be combined 
with the renormalization group equation for the running coupling and solved. In the late nineties in Ref. \cite{Blumlein:1997em,Diemoz:1987xu}
the evolution operator in the case of the running coupling has been constructed and fully analytical solutions of the non-singlet and singlet evolution equations at the next-to-next to leading order 
with small $x$ resummations included were found.  The recent developments of the solution for these first order differential equations may be found in  \cite{Ablinger:2018zwz}.  
At the NNLO  Mellin space solutions with the running coupling have been worked out in several numerical codes, for example \cite{Vogt:2004ns}, and various later numerical software 
packages may be found in the citations of \cite{Vogt:2004ns}. 
When these first order differential equations for the Mellin moments are solved, the usual way is to convert these moments back to  the Bjorken $x$-space
by making the inverse Mellin transformation which may be performed by evaluation of residues on the complex plane of the  Mellin moment 
\cite{Kosower:1997vj,Altarelli:2000dw,Ball:2005mj}. 
At the lowest order in the running coupling the calculation may be done analytically, however even at this level  a lot of work is required
when the real QCD case is considered instead of simple models.  There are different software packages available to do all these steps analytically, at least at the leading order. 
At higher orders new advanced analytical software tools exist. They are based on using concepts from algebraic geometry like a shuffle product \cite{Ablinger:2010kw,Ablinger:2013hcp}. 
Shuffle product is used in the construction of single-valued harmonic polylogarithms. Harmonic polylogarithms are described in  \cite{Remiddi:1999ew}.  
There are numerical packages which solve integro-differential equations as they are written without taking the Mellin moments, solving first order differential equations and then 
transforming the moments back to $x$ space. For example,  a numerical software package ``PartonEvolution'' has been developed in \cite{Weinzierl:2002mv}. 
Another numerical package QCDNUM has been created later \cite{Botje:2010ay,Botje:2016wbq}.

In the shadow of all these  achievements cited in the previous paragraphs our approach is an attempt to look differently at the contour integrals arising in solutions to the DGLAP equation. 
Here we propose an alternative way in which evaluation of the inverse Mellin transformation reduces to calculation of the inverse Laplace transformation of the Jacobian of the corresponding  complex map. 
These diffeomorphisms may be performed  in the complex plane of the  Mellin moment.  
These complex maps make the structure of the integrands uniform reducing it in many  of the cases to the standard tables like  \cite{gradshteyn-2015a}.  
Then, we may convert them by one more transformation to the Barnes integrals. 
These would allow their systematic classification in the terms of generalized hypergeometric functions. Any systematic classification is useful in construction of computer algorithms.

Due to the significant computational progress of the last decades (see for example Refs. \cite{Moch:2004pa,Vogt:2004mw,Botje:2010ay,Botje:2016wbq})  the perturbative solution to the DGLAP equation is already 
computed up to N$^2$LO for the Mellin moments of parton distribution functions with full inclusion of running coupling  and then the corresponding particle distribution functions  (quark, gluon and some other combinations)
were obtained numerically. 
However, approximate solutions 
to the DGLAP equation corresponding to simple models 
still have a practical value because they capture in a resumed way (in the sense of a compact expression) the behaviour of a given asymptotic regime. In particular, in the present paper it captures  
the Bessel-like behaviour with respect to square root of the product of logarithm on the Bjorken variable and logarithm of the momentum transfer  in the region of the small values of  $x$ 
when the main contribution comes from the gluon part  of the matrix DGLAP equation. On the other side, the approximate solutions should not be discarded because they serve as a consistency check for the current 
manipulations and formulas 
which lead us to the known obtained results. Another reason in favor of viability of such approximate solutions is  that in the low momentum transfer regime of the DGLAP equations the numerical 
solutions just start to show bad behaviour and  one may at least make some estimations in such limits by using these solutions and then obtain novel relations and interpretations.

In the next Section we consider the necessary formulas which may be found in the Gradshteyn and Ryzhik tables \cite{gradshteyn-2015a}. 
All the necessary formulas are related by integral transformations which are given explicitly.
We pay some attention to the relation between the Bessel function and the confluent function, and to different integral representations of the generalized hypergeometric functions. 
In Section 3 we convert the contour integral solution to the DGLAP equation to the Laplace transformation  of the complex Jacobian. The Jacobian corresponds to a complex map selected for a given transformation.
Finally, in Section 4 we make a transparent trick with help of which we re-write the Laplace transformation  of the complex Jacobians in a form of the Barnes integrals.

\section{Preliminary}

The only purpose of this Section is to collect together from Ref.  \cite{gradshteyn-2015a} all the formulae necessary for use in the next Sections. These formulae are not new, each of them is at least 
one hundred fifty years old, the same book of integral tables  \cite{gradshteyn-2015a} is quite old too. However, all the formulas that we have taken from \cite{gradshteyn-2015a} 
may be related by integral transformations from each one to another. We do all these transformations explicitly in this Section but it is probable that we are not first who publish these intermediate steps taking into account 
the age of  Ref. \cite{gradshteyn-2015a}. 

We start in Subsection 2.1 with the confluent function ${}_1 F_1.$ It is a particular case of generalized hypergeometric functions and may be written in terms of the Barnes integral. We start from this Barnes 
integral representation for ${}_1 F_1$ and obtain the corresponding series, then we obtain another integral representation for ${}_1 F_1,$ re-write this second integral representation for ${}_1 F_1$
in terms of the same series again and finally prove one useful relation
in terms of the same integral representation for ${}_1 F_1.$ In Subsection 2.2  two different integral representations of the Bessel function $I_0$ have been considered. Then, we show by a change of an integration variable 
that they are equivalent.   
The Bessel function $I_0$ is represented in terms of the confluent function  ${}_1 F_1.$  In Subsection 2.3 we consider the Barnes integral representation for the Gauss hypergeometric function  ${}_2 F_1$ 
and take this integral in terms of a series. Another integral representation is obtained for this Gauss hypergeometric function later. With help of this Euler integral representation for the Gauss hypergeometric function 
${}_2 F_1$ and with help of the established in Subsection 2.2 integral representation for the confluent function  ${}_1 F_1$ we reproduce the Laplace transform of the Bessel function $I_0$ in Subsection 2.4.

\subsection{Integral representations of hypergeometric function ${}_1 F_1$}

This subsection is dedicated to the generalized hypergeometric function ${}_1 F_1(a,c,x).$  Sometimes this function is called a confluent function. As a starting point to 
work with a hypergeometric function we use the Barnes integral representation for it 
\footnote{We omit the factor $1/2\pi i$  in front of each contour integral in the complex plane. The inverse factor is generated with the residues according to Cauchy integral formula.} 
\begin{eqnarray*}
{}_1 F_1(a,c,x) = \frac{\G(c)}{\G(a)}\oint_C~dz ~\frac{\G \le a + z \ri  \G \le -z \ri}{\G \le c + z \ri }(-x)^z ,
\end{eqnarray*}
which is basically a contour integral in the complex plane $z.$ The contour contains the vertical line which passes a bit to the left of the imaginary axis 
and should be closed to the right complex infinity in order to guarantee the vanishing of the contribution of the contour $C$ at the complex infinity. 
Due to this vanishing  the series which appears due to application  of the Cauchy integral formula will be convergent and turns out to be a traditional representation of the confluent function 
\begin{eqnarray} \label{series}
{}_1 F_1(a,c,x) = \sum_{k=0}^{\infty}\frac{(a)_k}{(c)_k}\frac{x^k}{k!}, 
\end{eqnarray}
where the Pochhammer symbol $(a)_k = \G(a+k)/\G(a).$   Eq.(\ref{series}) is  formula {\bf 9.210.1} of Ref. \cite{gradshteyn-2015a}. In such a case all the residues come form the Gamma function with the negative 
sign of its argument and we have 
the series  (\ref{series}) above. The Barnes integrals are  the contour integral representation of the generalized hypergeometric functions, however it is a convenient but not a unique integral representation 
of the hypergeometric functions  ${}_q F_p.$  There are several integral representations more. For example, in the case of   ${}_1 F_1(a,c,x)$ we may write 
\begin{eqnarray*}
{}_1 F_1(a,c,x) = \frac{\G(c)}{\G(a)}\oint_C~dz ~\frac{\G \le a + z \ri  \G \le -z \ri}{\G \le c + z \ri }(-x)^z  = \\
\frac{\G(c)}{\G(a)\G(c-a)}\oint_C~dz ~ {\rm B}(a+z,c-a) \G \le -z \ri (-x)^z = \\
\frac{\G(c)}{\G(a)\G(c-a)}\oint_C~dz ~ \int_0^1~d\tau~\tau^{a+z-1}(1-\tau)^{c-a-1} \G \le -z \ri (-x)^z = \\
\frac{\G(c)}{\G(a)\G(c-a)}\int_0^1~d\tau~\tau^{a-1}(1-\tau)^{c-a-1} \oint_C~dz\G \le -z \ri (-x\tau)^z = \\
\frac{\G(c)}{\G(a)\G(c-a)}\int_0^1~d\tau~\tau^{a-1}(1-\tau)^{c-a-1}e^{x\tau},
\end{eqnarray*}
and we may recognize formula  {\bf 9.211.2} of Ref. \cite{gradshteyn-2015a} for the confluent function. Of course, this representation turns out to be the series (\ref{series}) again, 
\begin{eqnarray*}
{}_1 F_1(a,c,x) = \frac{\G(c)}{\G(a)\G(c-a)}\int_0^1~d\tau~\tau^{a-1}(1-\tau)^{c-a-1}e^{x\tau} = \\
\frac{\G(c)}{\G(a)\G(c-a)}\sum_{k=0}^{\infty}\frac{x^k}{k!} \int_0^1~d\tau~\tau^{a+k-1}(1-\tau)^{c-a-1} = \frac{\G(c)}{\G(a)\G(c-a)}\sum_{k=0}^{\infty}\frac{x^k}{k!}{\rm B}(a+k,c-a) = \\
\frac{\G(c)}{\G(a)}\sum_{k=0}^{\infty}\frac{x^k}{k!}\frac{\G(a+k)}{\G(c+k)} = \sum_{k=0}^{\infty}\frac{(a)_k}{(c)_k}\frac{x^k}{k!}.
\end{eqnarray*}
This integral representation may be useful in calculation. For example, the following property may be proven 
\begin{eqnarray} \label{Gauss-4}
{}_1 F_1(a,c,x) = \frac{\G(c)}{\G(a)\G(c-a)}\int_0^1~d\tau~\tau^{a-1}(1-\tau)^{c-a-1}e^{x\tau} = \no\\
\frac{\G(c)}{\G(a)\G(c-a)}\int_0^1~d\tau~(1-\tau)^{a-1}\tau^{c-a-1}e^{x(1-\tau)} =   e^x{}_1 F_1(c-a,c,-x). 
\end{eqnarray}

\subsection{From the Bessel function $I_0$ to the hypergeometric function  ${}_1F_1$}

The traditional integral representation for the Bessel function $I_0$ turns out to be the well-known series in terms of the even powers of its argument, 
\begin{eqnarray*}
I_0(x) = \frac{1}{\G^2(1/2)}\int_{-1}^1~d\tau~(1-\tau^2)^{-\frac{1}{2}}e^{x\tau} =  \frac{1}{\G^2(1/2)}\sum_{k=0}^{\infty}\frac{x^{2k}}{(2k)!} \int_{-1}^1~d\tau~(1-\tau^2)^{-\frac{1}{2}}\tau^{2k} = \\
\frac{2}{\G^2(1/2)}\sum_{k=0}^{\infty}\frac{x^{2k}}{(2k)!} \int_0^1~d\tau~(1-\tau^2)^{-\frac{1}{2}}\tau^{2k} =  \frac{1}{\G^2(1/2)}\sum_{k=0}^{\infty}\frac{x^{2k}}{(2k)!} \int_0^1~dt~(1-t)^{-\frac{1}{2}}t^{k-\frac{1}{2}} = \\
 \frac{1}{\G^2(1/2)}\sum_{k=0}^{\infty}\frac{x^{2k}}{(2k)!}  {\rm B}\le\frac{1}{2},k+\frac{1}{2}\ri =  \\
 \frac{1}{\G(1/2)}\sum_{k=0}^{\infty}\frac{x^{2k}}{(2k)!k!}  \G\le k+\frac{1}{2} \ri = 
 \frac{1}{\G(1/2)}\sum_{k=0}^{\infty}\frac{x^{2k}}{(2k)!k!}  \frac{(2k)!}{4^k k!}   \G\le\frac{1}{2} \ri = 
\sum_{k=0}^{\infty}\frac{x^{2k}}{(k!)^2 4^k } =  \sum_{k=0}^{\infty}\frac{1}{(k!)^2}\le\frac{x^{2}}{4}\ri^k.
\end{eqnarray*}
This integral representation is {\bf 8.431.1} of Ref. \cite{gradshteyn-2015a}. This integral representation for the Bessel function $I_0$ is related to another integral representation which 
is  {\bf 9.211.2} of Ref. \cite{gradshteyn-2015a} by a change of the integration variable,
\begin{eqnarray} \label{second}
I_0(x) = \frac{1}{\G^2(1/2)}\int_{-1}^1~d\tau~(1-\tau^2)^{-\frac{1}{2}}e^{x\tau} =  \frac{2}{\G^2(1/2)}\int_{0}^1~dt~(1-(1-2t)^2)^{-\frac{1}{2}}e^{x(1-2t)} = \no\\
\frac{e^x}{\G^2(1/2)}\int_{0}^1~dt~(t(1-t))^{-\frac{1}{2}}e^{-2tx} = e^x{}_1 F_1\le\frac{1}{2},1,-2x\ri  = e^{-x}{}_1 F_1\le\frac{1}{2},1,2x\ri = \no\\
\frac{e^{-x}}{\G^2(1/2)}\int_{0}^1~dt~t^{-\frac{1}{2}}(1-t)^{-\frac{1}{2}}e^{2tx}.
\end{eqnarray}
This integral representation is {\bf 9.238.2} of Ref. \cite{gradshteyn-2015a}.
Thus, the integral representation (\ref{second}) allows us to re-write the Bessel function $I_0$ in terms of the confluent function ${}_1F_1$ and it will be more useful for us 
in Subsection 2.4 where we calculate the Laplace transform of the Bessel function $I_0.$

\subsection{Integral representations of hypergeometric function ${}_2F_1$}

In order to study the Gauss hypergeometric function ${}_2F_1$ we again use as the starting point the representation in terms of the Barnes contour integral, 
which may be re-written in a form of the well-known series for the Gauss hypergeometric function, Eqs. {\bf 9.100}  and  of {\bf 9.113} Ref. \cite{gradshteyn-2015a}.
\begin{eqnarray*}
{}_2 F_1(a,b;c,x) = \frac{\G(c)}{\G(a)\G(b)}\oint_C~dz ~\frac{\G \le a + z \ri \G \le b + z \ri   \G \le -z \ri}{\G \le c + z \ri }(-x)^z  = \sum_{k=0}^{\infty}\frac{(a)_k {(b)}_k}{(c)_k}\frac{x^k}{k!},
\end{eqnarray*}
Again, we have to take this contour containing the vertical line situated a bit to the left from the imaginary axis and close it in a such way that the complex infinity does not contribute. Where to close it 
depends on the absolute value of the variable $x.$ We suppose that $0<x<1$ and in this range we have to close it to the right hand side.  The residues come from the Gamma function with the negative sign 
of its argument and we obtain the series above. 

The Barnes integrals are  the contour integral representation of the hypergeometric functions,  it is not a unique integral representation of the hypergeometric 
functions  ${}_q F_p,$  there are other integral representations. In the case of   ${}_2F_1(a,b;c,x)$ we may transform
\begin{eqnarray}\label{Gauss}
{}_2 F_1(a,b;c,x) = \frac{\G(c)}{\G(a)\G(b)}\oint_C~dz ~\frac{\G \le a + z \ri \G \le b + z \ri   \G \le -z \ri}{\G \le c + z \ri }(-x)^z  = \no\\
\frac{\G(c)}{\G(a)\G(b)\G(c-b)}\oint_C~dz ~ {\rm B}(b+z,c-b) \G(a+z)\G \le -z \ri (-x)^z = \no\\
\frac{\G(c)}{\G(a)\G(b)\G(c-b)}\oint_C~dz ~ \int_0^1~d\tau~\tau^{b+z-1}(1-\tau)^{c-b-1}    \G(a+z) \G \le -z \ri (-x)^z = \no\\
\frac{\G(c)}{\G(a)\G(b)\G(c-b)}\int_0^1~d\tau~\tau^{b-1}(1-\tau)^{c-b-1} \oint_C~dz \G(a+z)  \G \le -z \ri (-x\tau)^z = \no\\
\frac{\G(c)}{\G(b)\G(c-b)}\int_0^1~d\tau~\tau^{b-1}(1-\tau)^{c-b-1} (1-x\tau)^{-a}.
\end{eqnarray}
This is formula  {\bf 9.111} of Ref. \cite{gradshteyn-2015a}  for ${}_2F_1.$  In the next Subsection we will use the integral representation (\ref{Gauss}) of the Gauss hypergeometric function 
in order to calculate the Laplace transform of the Bessel function.  In complete analogy to Eq. (\ref{Gauss-4}) we may prove identity  {\bf 9.131} of Ref. \cite{gradshteyn-2015a}  
\begin{eqnarray} \label{Gauss-3}
{}_2 F_1(a,b;c,x) = \frac{\G(c)}{\G(b)\G(c-b)}\int_0^1~d\tau~\tau^{b-1}(1-\tau)^{c-b-1} (1-x\tau)^{-a} = \no\\ 
\frac{\G(c)}{\G(b)\G(c-b)}\int_0^1~d\tau~(1-\tau)^{b-1}\tau^{c-b-1} (1 -x + x\tau)^{-a} = \no\\
\frac{\G(c)}{\G(b)\G(c-b)}(1-x)^{-a}\int_0^1~d\tau~\tau^{c-b-1} (1-\tau)^{b-1}\le 1 - \frac{x}{x-1}\tau\ri^{-a} = \no\\
(1-x)^{-a} {}_2 F_1 \le a,c-b;c,\frac{x}{x-1}\ri.
\end{eqnarray}

\subsection{The Laplace transform of the Bessel function $I_0$}

In this Subsection we reproduce a result for the Laplace transform of the Bessel function $I_0.$  The result is written in {\bf 17.13.109} of Ref. \cite{gradshteyn-2015a}. 
When we do this transformation, we suppose that $z$ is in the corresponding domain of the complex plane,  that is, on the right hand side of the critical exponent of the Bessel function \cite{Alvarez:2016juq}.  
\begin{eqnarray*} 
\int_0^{\infty}I_0(x)e^{-xz}dx = \frac{1}{\G^2(1/2)}\int_{0}^1~dt~t^{-\frac{1}{2}}(1-t)^{-\frac{1}{2}}\int_0^{\infty}e^{-x}e^{2tx} e^{-xz}dx = \no\\
\frac{1}{\G^2(1/2)}\int_{0}^1~dt~t^{-\frac{1}{2}}(1-t)^{-\frac{1}{2}}\frac{1}{1-2t+z} = \no\\ 
\end{eqnarray*}
\begin{eqnarray} \label{dir}
\frac{1}{z+1}\frac{1}{\G^2(1/2)}\int_{0}^1~dt~t^{-\frac{1}{2}}(1-t)^{-\frac{1}{2}}\le1-\frac{2}{z+1}t\ri^{-1} = \no\\
\frac{1}{z+1}{}_2 F_1\le 1,\frac{1}{2};1,\frac{2}{z+1}\ri = \frac{1}{z+1} \frac{1}{\G(1/2)}\oint_C~dz ~\frac{\G \le 1 + z \ri \G \le \frac{1}{2} + z \ri   \G \le -z \ri}{\G \le 1 + z \ri }\le-\frac{2}{z+1} \ri^z  = \no\\
\frac{1}{z+1} \frac{1}{\G(1/2)}\oint_C~dz ~\G \le \frac{1}{2} + z \ri   \G \le -z \ri \le-\frac{2}{z+1} \ri^z =  \no\\
\frac{1}{z+1} \le 1 -\frac{2}{z+1} \ri^{-\frac{1}{2}}  = \frac{1}{z+1} \le \frac{z+1}{z-1} \ri^{\frac{1}{2}} = \frac{1}{\sqrt{z^2-1}}
\end{eqnarray}
This inverse square root is very known result for the Laplace transform of the Bessel function $I_0$ and may be found in many tables of integrals.

\section{The DGLAP contour integral solution as Laplace transform of the complex Jacobian}

This is the main Section. Everything we have written in the previous Sections was a preparation for this Section.  Here we calculate the contour integrals of this type 
\begin{eqnarray} \label{type}
\phi(x,u) =  \int_{-1+\delta-i\infty}^{-1+\delta+i\infty}~dN \frac{x^{-N}}{N+1} u^{ \displaystyle{1/(N+1)}} 
\end{eqnarray}
by making complex maps in the plane of the Mellin moment. Here $x \in [0,1]$ and $u \in [0,\infty[$ are external variables. This contour integral represents solution to 
the DGLAP integro-differential equation which plays an important role in Quantum Chromodynamics. We commented on this equation in the Introduction.  The traditional way is to calculate this contour 
integral directly by evaluating residues according to the Cauchy integral formula,  
\begin{eqnarray*} 
\phi(x,u) = \int_{-1+\delta-i\infty}^{-1+\delta+i\infty}~dN \frac{x^{-N}}{N+1} u^{ {1/(N+1) }} = 
\end{eqnarray*}
\begin{eqnarray*}
x\sum_{k=0}^{\infty}\frac{1}{k!}\le\ln{u}\ri^k \int_{-1+\delta-i\infty}^{-1+\delta+i\infty}~dN \frac{x^{-N-1}}{(N+1)^{k+1}} \no \\
= x\sum_{k=0}^{\infty}\frac{1}{k!}\le\ln{u}\ri^k \frac{(-\ln{x})^{k}}{k!} = xI_0\le 2\sqrt{\ln{u}\ln{\frac{1}{x}}} \ri.
\end{eqnarray*}
In this example this is the shortest way of getting the result for this integral. However, we would like to reproduce this result by using complex geometry in order  to make a map in the complex plane 
of the Mellin moment.  For this purpose  we re-write the previous integral 
\begin{eqnarray} \label{line} 
\phi(x,u) =  \int_{-1+\delta-i\infty}^{-1+\delta+i\infty}~dN \frac{x^{-N}}{N+1} u^{ \displaystyle{1/(N+1)}}  = \int_{-1+\delta-i\infty}^{-1+\delta+i\infty}~dN \frac{1}{N+1}e^{-N\ln{x} + \ln{u}/(N+1)}
\end{eqnarray}
and choose a new complex variable $M$ of integration in such a way that 
\begin{eqnarray} \label{map} 
M\sqrt{ \ln{u}\ln{\frac{1}{x}}  } = -N\ln{x} + \frac{\ln{u}}{N+1}.  
\end{eqnarray}
This relation defines $M$ as a function of the initial complex variable $N$ \cite{Kondrashuk:2019cwi}.    Introducing for the brevity a notation 
\begin{eqnarray*} 
w \dis{\equiv \sqrt{ \frac{\ln{u}}{\ln{\frac{1}{x}}}}},
\end{eqnarray*} 
we may write 
\begin{eqnarray*} 
M = \frac{N}{w} + \frac{w}{N+1},  
\end{eqnarray*} 
from which it follows that in the inverse mapping from $M$ to $N$ the initial variable $N$ must satisfy the quadratic equation 
\begin{eqnarray*} 
N^2 + (1-wM)N + (w^2 -Mw) = 0,
\end{eqnarray*} 
which has two roots 
\begin{eqnarray}\label{map-inv} 
N^{(+)} = \frac{ Mw - 1 + \sqrt{(Mw+1)^2 - 4w^2}}{2}, ~~~ N^{(-)} = \frac{ Mw - 1 - \sqrt{(Mw + 1)^2 - 4w^2}}{2}.
\end{eqnarray} 
If we do not want to change the contour orientation while mapping  from $N$ to $M$ according to Eq. (\ref{map}) we should use a positive branch of the map (\ref{map-inv}). Thus, we have chosen the map 
\begin{eqnarray}\label{map-inv-2} 
N(M) = N^{(+)}(M) = \frac{ Mw - 1 + \sqrt{(Mw + 1)^2 - 4w^2}}{2}.
\end{eqnarray} 
We could choose the negative branch too, but in this case we should change the sign of the integral because the  contour orientation would be changed under this map. 

Having the complex map chosen in Eq. (\ref{map-inv-2}) we may re-write the integrand in terms of the new complex variable $M.$  From Eq. (\ref{map}) we have   
\begin{eqnarray*} 
dM =  \le \frac{1}{w} - \frac{w}{(N+1)^2} \ri dN ~~~ \Rightarrow ~~~ \frac{dN}{dM} \frac{1}{N+1} =  \le\frac{N+1}{w} - \frac{w}{N+1}\ri^{-1} = \frac{1}{\sqrt{\le M+{1}/{w}\ri^2 - 4}}. 
\end{eqnarray*} 
Now we may proceed the line (\ref{line}) in terms of the new integration variable $M$  
\begin{eqnarray} \label{line-2} 
\phi(x,u) =  \int_{-1+\delta-i\infty}^{-1+\delta+i\infty}~dN \frac{x^{-N}}{N+1} u^{ \displaystyle{1/(N+1)}}  = \int_{-1+\delta-i\infty}^{-1+\delta+i\infty}~dN \frac{1}{N+1}e^{-N\ln{x} + \ln{u}/(N+1)} = \no\\
\oint_C dM  \frac{e^{M\sqrt{ \ln{u}\ln{\frac{1}{x}}}}  }{\sqrt{\le M+{1}/{w}\ri^2 - 4}}   =  e^{-\sqrt{\ln{u}\ln{\frac{1}{x}}}/w} \oint_{C'} dM  \frac{e^{M\sqrt{ \ln{u}\ln{\frac{1}{x}}}}  }{\sqrt{ M^2 - 4}} = \no\\
e^{-\ln{\frac{1}{x}}} \oint_{C''} dM  \frac{e^{2M\sqrt{ \ln{u}\ln{\frac{1}{x}}}}  }{\sqrt{ M^2 - 1}} = xI_0\le 2\sqrt{\ln{u}\ln{\frac{1}{x}}} \ri.
\end{eqnarray}
Under this sequence of the maps, the contour in the complex plane changes its shape from the vertical line parallel to the imaginary axis to a very complicated form $C''$ at the end of this chain of transformations. 
We wrote the last equality because we know that $C''$ may be transformed to the vertical line parallel to the imaginary axis in the complex plane $M.$ In such a way we may write 
the last equality basing on the Laplace transform of the Bessel function found in Eq. (\ref{dir}). The direct proof of the last equality will be given in the next Section by explicit calculation
of the integral. 

In general, our purpose is to obtain the result for the contour integrals that represent solution to  DGLAP integro-differential equation by making 
complex diffeomorphism  in the plane of the Mellin moment $N$ related to the Bjorken variable $x.$  Why do we do complex diffeomorphisms and represent the contour integral of this type (\ref{type}) as 
a Laplace transformation of a Jacobian? We find this way more systematic in order to classify the obtained results in terms of special functions. 
One reason is that standard table integrals may appear as it happened in the case of Eq.(\ref{line-2}). Another reason is 
that the Laplace transformation may be represented in a form of the Barnes integrals (we consider this second reason in detail in the next section).  
These two reasons are fundamental in construction of computer algorithm \cite{Kondrashuk:2019cwi} in the real world case of QCD in which the functions of anomalous dimension are highly complicated and 
contain many terms. At the three loop level these functions can be found in Refs.  \cite{Moch:2004pa,Vogt:2004mw}.

The suggested solution to the DGLAP equation given in Eqs. (\ref{type}) and (\ref{line-2}) corresponds to the kinematic region of small $x$  in which this approximate solution makes a sense. Strictly speaking, we 
need to have as the result a function singular at $x=0.$ This corresponds to the singularity at a point $N=1$ in the complex $N$ plane. The terms in the matrix of anomalous dimensions singular at this point correspond 
to the dominance of the 
gluon distribution function in the small $x$ region. The matrix form may be reduced to the DGLAP equation for the dominant parton distribution in the region of the small $x.$ This would be a quite good approximation 
to the matrix DGLAP integro-differential equation.  To have such a behaviour we would do a small modification and take another model for the anomalous dimension $\gamma(N)$ which is singular at the point $N=1.$
In general, the kinematic region of the DGLAP equation is given by the Bjorken limit in which the ratio of the momentum transfer to the Bjorken $x$ is large. 
By increasing or decreasing the momentum transfer $u$ one has to increase or decrease the $\phi(x,u)$ distribution by a double logarithm law found in Eqs. (\ref{type}) and (\ref{line-2}) as a result of  the pure gluonic 
DGLAP dynamics.  

As we have mentioned in the Introduction, the practical value of this simple model is that its solution to the DGLAP equation may be used to estimate qualitatively the behaviour of the dominant distribution function 
nearby $x=0,$ that is for the very small $x.$  However, this model may appear to be even more useful than a simple approximation for the dominant distribution or a consistency check for the numerical or analytical 
calculation based on a powerful software are. As it is known, there are several groups in the world which make global analysis of the parton distribution functions taking into account recent data from the LHC 
\cite{Hou:2019efy,Dulat:2015mca,nnpdf-1,nnpdf-2,Ethier:2020way}. These analysis allows them  to fix many parameters of initial parton distribution functions from data only because  they cannot be computed from first principles.
Despite such a fitting procedure is rather arbitrary on how to choose the form of parton distribution function as well as number of free parameters, they tend to be some combination of Euler beta functions, 
which parametrize the parton distribution  at some scale \cite{Ball:2016spl,Alekhin:2003ev,Blumlein:2006ws,Alekhin:2012ig} and than evolve according to the DGLAP integro-differential equation. 
Due to big amount of data the software for the fitting of the  PDF parameters and for the PDF evolution is created  on the principles of neural networks \cite{nnpdf-1,nnpdf-2}.
This simple model may be used to train the neural networks.\footnote{I.K. is grateful to Andrei Arbuzov for attracting attention to this point.} We also note that it is not a unique simple model which may 
be used for this purpose. We may create several simple models 
that solve the DGLAP integro-differential equation with different splitting functions by adjusting the shape functions at some given scale in order to combine them with Jacobians of complex diffeomorphisms in the plane of 
the Mellin moment variable in easily integrable factors.

\section{From inverse Laplace transformations of Jacobians to the Barnes integrals}

In the previous Section we have done complex diffeomorphisms in order to represent the initial form (\ref{type}) of the solution to the DGLAP equation  in terms of contour integral to the Jacobian form (\ref{line-2}) 
of this solution.  This Jacobian form may be a standard integral from the well-known integral tables \cite{gradshteyn-2015a}, for example the Bessel function $I_0$ in the case of our model. 
We have proved that this is Bessel function by the direct Laplace transformation in Subsection 2.4. Now we evaluate the inverse Laplace transformation,
\begin{eqnarray} \label{Fubini}  
 \oint_{C''} dz  \frac{e^{zx}  }{\sqrt{ z^2 - 1}} =  \frac{1}{(-1)^{1/2}} \frac{1}{\G(1/2)}\oint_{C''}dz~e^{zx}\oint_{C}~du ~\G \le u + \frac{1}{2} \ri  \G \le -u\ri \le-z^2\ri^u.
\end{eqnarray}
At this moment we change the order of integration and we have 
\begin{eqnarray} 
\int_0^{\infty}x^{-2u-1}e^{-xz}dx = z^{2u}\G(-2u) ~~~ \Rightarrow ~~~ \oint_{C''}dz~e^{zx}z^{2u} = \frac{x^{-2u-1}}{\G(-2u)}
\end{eqnarray}
In such a way we may say that we have Hankel contour $C''.$ Now we may proceed line $(\ref{Fubini})$ 
\begin{eqnarray*}   
 \oint_{C''} dz  \frac{e^{zx}  }{\sqrt{ z^2 - 1}} =  \frac{1}{(-1)^{1/2}} \frac{1}{\G(1/2)}\oint_{C}du~ \G \le u + \frac{1}{2} \ri  \G \le -u\ri \frac{x^{-2u-1}(-1)^u}{\G(-2u)} =\\
\frac{1}{(-1)^{1/2}} \frac{1}{\G(1/2)}\oint_{C}du~ \G \le u + \frac{1}{2} \ri  \G \le -u\ri \frac{x^{-2u-1}(-1)^u2^{1+2u}\G\le 1/2 \ri}{\G(-u)\G\le -u + 1/2 \ri } = \\
\frac{1}{(-1)^{1/2}} \frac{1}{\G(1/2)}  \oint_{C}du~ \G \le u + \frac{1}{2} \ri  \G \le -u\ri \frac{x^{-2u-1}(-1)^u2^{1+2u}\G\le 1/2\ri}{\G(-u)\G\le -u + 1/2  \ri } = \\
\frac{1}{(-1)^{1/2}} \le\frac{2}{x}\ri \oint_{C}du~ \frac{\G \le u + 1/2 \ri}{\G\le -u + 1/2 \ri}  \le-\frac{4}{x^2}\ri^u  =  
\le\frac{2}{x}\ri^2 \oint_{C_1}du~ \frac{\G \le u + 1 \ri}{\G\le -u \ri}  \le-\frac{4}{x^2}\ri^u = \\
\le\frac{2}{x}\ri^2 \oint_{C_2}du~ \frac{\G \le 1 - u\ri}{\G\le u \ri}  \le-\frac{x^2}{4}\ri^u = \le\frac{2}{x}\ri^2\sum_{k=1}^{\infty}\frac{(-1)^k}{\le(k-1)!\ri^2}  \le-\frac{x^2}{4}\ri^k = 
\sum_{k=0}^{\infty}\frac{1}{\le k!\ri^2}  \le\frac{x^2}{4}\ri^k = I_0(x).
\end{eqnarray*}
By comparing this equation and Eq.(\ref{line-2}) we may write this identity  
\begin{eqnarray*}   
\oint_{C''} dM  \frac{e^{Mx}}{\sqrt{ M^2 - 1}}   =   \le\frac{2}{x}\ri^2 \oint_{C_2}du~ \frac{\G \le 1 - u\ri}{\G\le u \ri}  \le-\frac{x^2}{4}\ri^u,
\end{eqnarray*}
in which on the left hand side we have the Jacobian form (\ref{line-2}) of the contour integral   (\ref{type}) and on the right hand side we have the Barnes integral. 
We may compare these two different forms and observe that the Barnes integral form is more useful in order to classify the result in terms of the special functions.  
Roughly speaking, in order to write the Bessel function on the left hand side it is better to keep  Gradshteyn and Ryzhik tables in hands and on the right hand side we do not 
need them. The second reason to represent the initial contour integral  (\ref{type}) in terms of the Barnes integrals is 
to have a uniform well-studied representation for all the contour integrals involved in the calculation. The third reason is that the Jacobian form on the 
left hand side will frequently contain multivalued functions, in our case it is a square root in the denominator. We avoid the integration over cuts which is necessary in the 
complex plane of the integration variable $M$ in the Jacobian form. We do not have any cuts in the Barnes integral representation on the right hand side.  
The fourth reason to prefer the result written in the form of the Barnes integrals is that the Jacobians will be more and more complicated in higher orders while 
the contour integrals of ratios of some products of several Gamma functions are  well-studied, their structure is more uniform and better analyzable. Ratios of Gamma functions 
in the integrands of the contour integrals appear frequently in quantum field theory calculation \cite{Smirnov, Allendes:2009bd,Allendes:2012mr,Gonzalez:2012gu,Gonzalez:2012wk,Kniehl:2013dma,
Gonzalez:2016pgx,Gonzalez:2018gch}.

\section{Conclusion}

In the previous paper \cite{Kondrashuk:2019cwi} we have communicated that the BFKL equation \cite{Lipatov:1976zz,Fadin:1975cb,Kuraev:1976ge,Kuraev:1977fs,Balitsky:1978ic} may be obtained from the DGLAP 
equation via a complex map in the domain of the 
contour integral (\ref{type}) that represents the solution to this DGLAP integro-differential equation. This is why the BFKL integro-differential equation may be considered as a dual 
equation to the DGLAP integro-differential equation. The question of duality between the BFKL equation and the DGLAP equation has been arisen in \cite{Altarelli:2000dw}.  
Also, in Ref. \cite{Kondrashuk:2019cwi} we have proposed that a complex map in the plane of the Mellin moment may be used to transform the solution to the DGLAP integro-differential equation 
obtained in the form of contour integral of the type (\ref{type}) to the form which represents the Laplacian transformation of the Jacobian of this complex map. In the present paper 
we have shown that the Jacobian form of the contour integral may be further transformed to the Barnes contour integrals. The Barnes integrals are contour integrals in which the integrand is a ratio of 
products of several Gamma functions and the result of their integration may be written in terms of special functions in a traditional well-studied way. The representation in terms of the 
Barnes integrals is useful in construction of computational algorithms \cite{Kondrashuk:2019cwi}.

\subsection*{Acknowledgments}

The work of G.A. was supported in part by the joint DAAD/CONICYT scholarship 2015/57144001. 
The work of I.K. was supported in part by Fondecyt (Chile) Grants Nos. 1040368, 1050512 and 1121030, by DIUBB (Chile) Grant Nos.  125009,  GI 153209/C  and GI 152606/VC.  
These results were presented in the talk of I.K. at XXXI Jornada Matem\'atica de la Zona Sur, Valdivia, Chile, Abril 25 - 27, 2018.
He is grateful to Francisco Correa and Felipe van Diejen for inviting him to give a talk at Section ``Mathematical Physics'' of this annual scientific meeting.

\end{document}